# Deep learning based low-dose synchrotron radiation CT reconstruction

*Ling Li*[1,2], *Yu Hu*[1, *]

[1]Institute of High Energy Physics, CAS, 100049 Beijing, China
[2]University of Chinese Academy of Sciences, 100049 Beijing, China

**Abstract.** Synchrotron radiation sources are widely used in various fields, among which computed tomography (CT) is one of the most important. The amount of effort expended by the operator varies depending on the subject. If the number of angles needed to be used can be greatly reduced under the condition of similar imaging effects, the working time and workload of the experimentalists will be greatly reduced. However, decreasing the sampling angle can produce serious artifacts and blur the details. We try to use a deep learning model which can build high quality reconstruction sparse data sampling from the angle of the image and ResAttUnet are put forward. ResAttUnet is roughly a symmetrical U-shaped network that incorporates similar mechanisms to ResNet and attention. In addition, the mixed precision is adopted to reduce the demand for video memory of the model and training time.

## 1 Introduction

Synchrotron radiation is electromagnetic radiation emitted by charged particles moving at varying speeds in a magnetic field near the speed of light. Synchrotron radiation source is widely used in atomic physics, molecular environment science, radiation therapy, nanoscience, nuclear physics, medicine, medical diagnosis and other fields because of its wide spectrum distribution, high brightness and high collimation. The main experimental methods of synchrotron radiation are also varied, including X-ray lithography, high spatial resolution X-ray imaging, X-ray diffraction and scattering, etc. In practical application, some experiments are very time-consuming, and some experiments require high operating accuracy such as computed tomography (CT). If the number of angles needed to be used can be greatly reduced under the condition of similar imaging effects, the working time and workload of the experimentalists will be greatly reduced.

[*] Corresponding author: huyu@ihep.ac.cn

Analytical algorithms and iterative algorithms are two main types of traditional CT reconstruction methods. The theoretical cornerstone of the analytic algorithms is Radon Transform[1]. One of the most widely used analytical algorithms is the filtering back projection reconstruction algorithm (FBP)[1]. If the number of sampling angles is reduced, the angle between the two pauses will also increase, and the reliability of the data in the interpolation region will be reduced, so the error caused by interpolation will be larger. The iterative algorithms are based on the assumption of discretization. Set the discrete points on the plane as unknowns and the equations are set up by using the rays passing through the object. To reconstruct the image is essentially to solve the equations. The core of iterative algorithms is to use gradient descent method to reduce the error until the solution of the equations is iterated. Because of the large amount of data, overdetermined or underdetermined, ill-conditioned problems, the iterative method is often used instead of the traditional method to solve the equations.

The high similarity between iterative algorithms and neural network makes it possible for neural network to be applicable to the realization of CT reconstruction conjecture. Many researchers have attempted to use deep learning network to post-process CT images and achieved excellent results[3, 4]. The landmark network architecture in this field is Unet[2]. Unet is often used in CT image segmentation. In this paper, the author tries to add a structure similar to ResNet[5] on the basis of Unet network structure to reduce the influence of gradient disappearance of neural network, so that the network can improve its performance with the deepening of depth. At the same time, the attention mechanism[6,7] is also added to the network, which can make the network reconstruction pay more attention to the subtle structure of the image through the weighted algorithm. These structures are often the focus of the experimenter's attention, such as fine lines or gaps in an object's interior. Because of the intensity and high brightness of synchrotron radiation source, its CT data is often much larger than the general medical CT data. However, the neural network training process has a large demand for video memory, so the mixed precision training technique is adopted in this paper. The default data type of neural network is FP32. Using FP16 not only reduces video memory usage, but also speeds up training. However, directly replacing FP32 with FP16 will cause grad overflow/underflow and rounding error. In mixed precision training, FP16 is used for storage and multiplication, FP32 is used for accumulation, and the idea of Loss Scaling is used to minimize the error caused by using FP16. In the experimental results section of the fourth chapter, it is shown that the use of mixed precision training reduces video memory occupation and training time.

## 2 Synchrotron Radiation CT Reconstruction Method Based on Neural Network

The network ResAttUnet proposed in this paper is relatively symmetric on the whole, and it cannot complete the transformation from sinusoidal domain to plane domain, so it cannot directly reconstruct the original data into images. We first need to use FBP to convert the original data from the sinusoidal domain to the plane domain, and then use the neural network to remove artifacts, refine the image and other work.

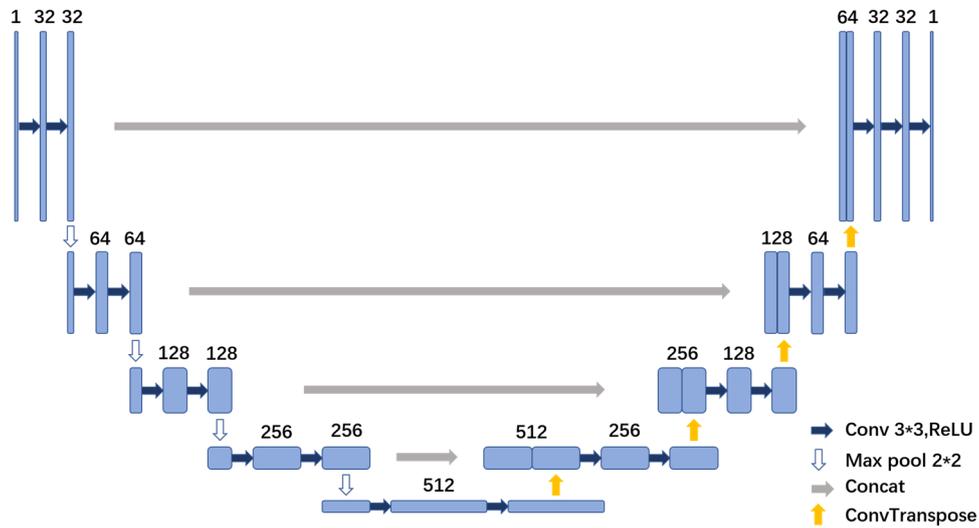

**Fig. 1.** Unet model in the artical

Unet network was proposed in 2015. Once it was proposed, it swept the field of CT medical images and became a landmark neural network structure. The network as a whole presents a symmetrical U-shape, which gives the input and output data the same dimension. On the left side of the U-shaped structure, with the deepening of the depth of the network, the area perceived by the filter is constantly enlarged, and the field of vision is constantly enlarged. The output image can also integrate features of different scales. The right side of the U shape uses convtranspose to make the output image bigger. Images of the same depth on the left and right sides of U-type are connected by skip connection. This structure can make the data on the right side of U-type supplement the relatively original information on the left side. Since the current detector pixel has been improved, generally reaching 2K, and more than 10K detectors will be used in the future, which will increase the memory, so the parameters of the Unet model we used as the control will be half of the original.

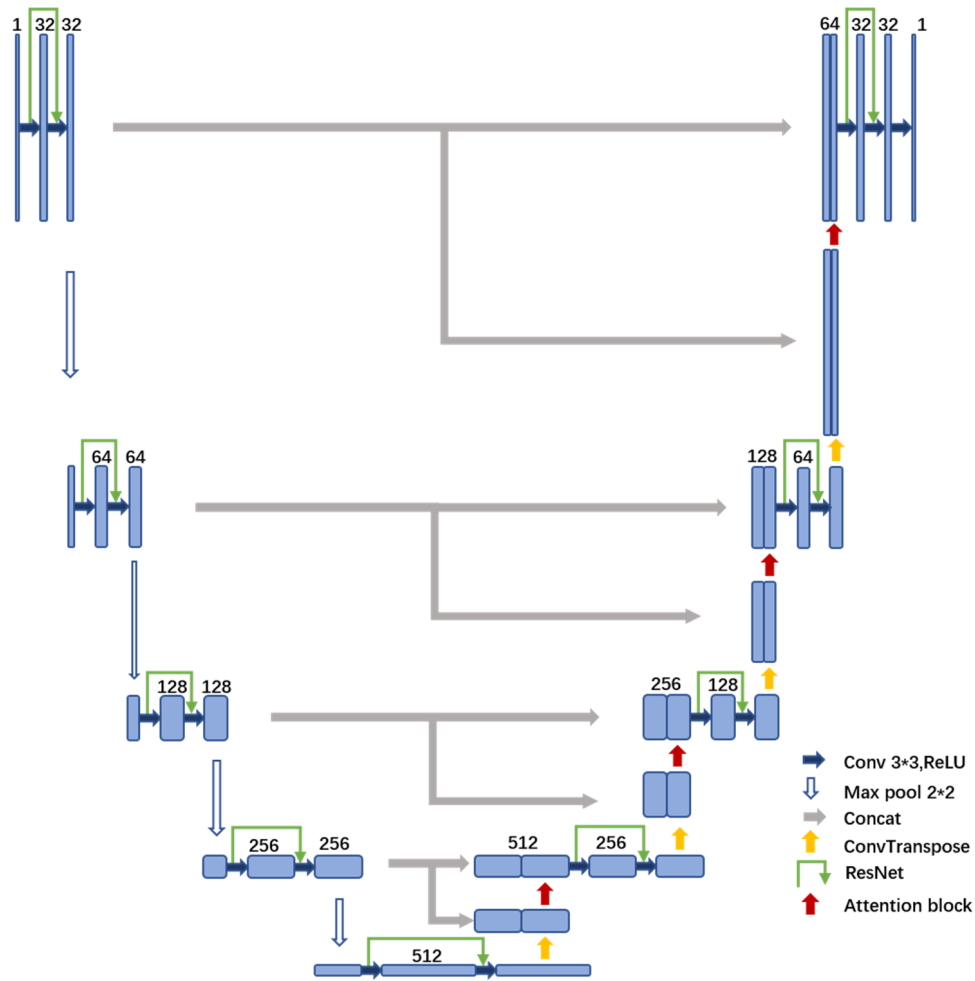

**Fig. 2.** The architecture of ResAttUnet

The ResAttUnet structure of the model proposed in this paper is shown in the **Fig. 2** above. Its overall framework is similar to the Unet, but on the basis of the Unet it joins the two structures, ResNet-like structure and attention mechanism. Both the left and right sides of the U-shape have a structure similar to ResNet. ResNet is proposed to alleviate the gradient vanishing problem of deep neural networks. With the number of layers of neural network increases, its effect should not be poorer compared to the network with shallow layer. But in experiments, deeper neural networks may tend to have larger errors, which is largely caused by the disappearance of the gradient of neural network. ResNet connects the neural network layer near the input to the layer near the output. If the output of a layer is already a good fit for the desired result, adding another layer will not make the model worse.

The ResNet-like structural data used in this article skips only one conv, and its output is actually the output of the second CONV plus the input. Their specific differences are shown in **Fig. 3**.

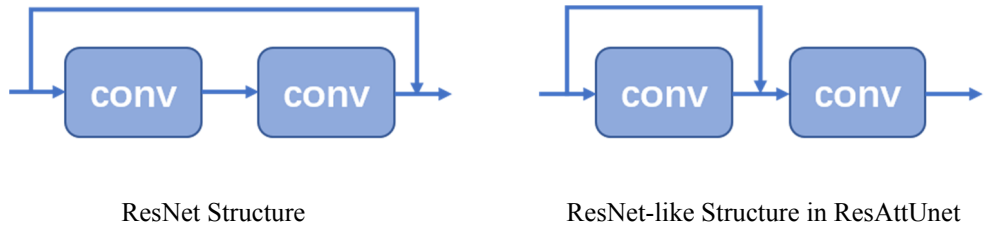

ResNet Structure                    ResNet-like Structure in ResAttUnet

**Fig. 3.** The contrast between ResNet and ResNet-like Structure in ResAttUnet

Attention mechanism can make the model ignore irrelevant information and pay attention to important information. The attention mechanism, so that the network can better deal with the internal details of the reconstructed image. As is shown in **Fig. 4**, the data on the left and right sides of the U-shaped structure into two 1 * 1 convolutions are input respectively, and then input a 1 * 1 convolution after concatenating the result. The output filter of this convolution is 1. Finally, the data on the right side of the U-shaped structure is multiplied by the convolution weight to get the output.

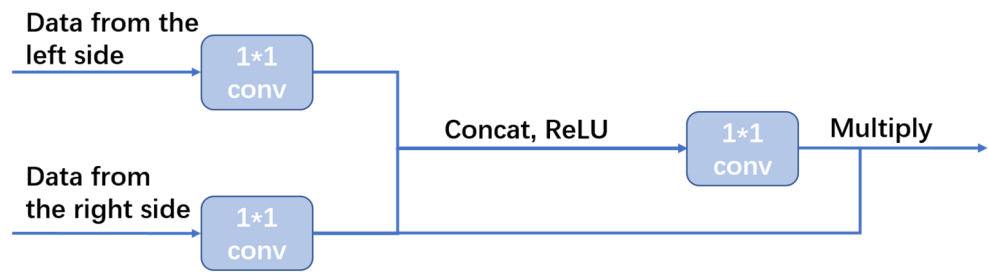

**Fig. 4.** The structure of attention block

## 3 Experiment

### 3.1 data set introduction

First, the simulated tomographic data is used to train the proposed network in the training process. We used the randomly generated foam phantoms 3D objects. To generate a foam phantom, non-overlapping randomly-placed spheres with randomly intense and varying sizes were removed from a cylinder of a single material. Tomographic projections of the objects were simulated using TomoPhantom[8]. In addition, we have a set of real data for testing. Our actual data source is Beijing Synchrotron Radiation Facility (BSRF) [9].

Sparse reconstructions of real data and simulated data are needed, and the subsampling multiple used in this experiment is 20 times, that is, one out of every 20 angles of the original data is selected as the data for reconstruction. There are obvious artifacts in sparse reconstruction images. The sparse reconstructed images of the simulated data are used as the training set and validation set, and the real value of the simulated data are used as the training set label and validation set label.

### 3.2 Conducting experiments

After the data set is read in, the data is uniformly processed so that each pixel is represented by 8-bit binary. The scale of simulated data and real data as well as their image after sparse reconstruction are unified to prevent model training failure due to inconsistency of data scale.

The next step consists in building the model. The final number of filters in each layer on the left side of the U type is 32, 64, 128, 256 and 512 respectively. The number of filters in each layer on the right side of the U corresponds to that on the left side of the U. The parameters of the Attention block are determined by the number of filters in the same layer on the left of the U and the number of filters in the conv block below it.

Our criterion selects MSELoss and the optimizer selects Adam optimizer. Since mixed accuracy training is adopted, "eps= 1e-7" need to be declared when setting the Adam optimizer. The initial learning rate is 3e-4, and as the epoch increased, the learning rate decreased dynamically. For every 10 epochs, the learning rate becomes 0.95 of the original size. Mixed precision training can be set to the level, O0 for pure FP32 training, O3 for pure FP16 training. From O1 to O3, fp16 accounts for more and more. We set the parameter OPT_LEVEL for mixed precision training as "O2", that is, except for the BATCH norm, almost all calculations are performed using FP16. Set the parameter OPT_LEVEL ="O2" for mixed precision training. The epoch number for training is 65.

In addition, we also used the Unet model and the ResAttUnet model which did not use mixed precision training as a comparison. In the process of implementing the Unet model as a control experiment, except the neural network structure is different, the processing methods and parameters of other parts are exactly the same. In the ResAttUnet model without mixed precision training, except the parameter OPT_LEVEL ="O0" for mixed precision training, the processing methods of other parts are exactly the same.

## 4 Experimental results and analysis

### 4.1 Evaluation criteria

Our evaluation criteria mainly include two indicators: PSNR (Peak Signal-to-Noise Ratio) and SSIM (Structural Similarity). These two indexes are often used for image evaluation. Since MSE (Mean Squared Error) is the mean energy difference between the real image and the noisy image, and the difference is the noise. PSNR is the ratio of the peak signal energy and MSE. PSNR is defined as follows:

$$PSNR = 10log_{10}\frac{MaxValue^2}{MSE} = 10log_{10}\frac{2^{bits}-1}{MSE} \quad (1)$$

The SSIM formula is based on three comparative measures between samples x and y: luminance, contrast, and structure. Its final definition is as follows.

$$SSIM(x,y) = \frac{(2\mu_x\mu_y+c_1)(2\sigma_{xy}+c_2)}{(\mu_x^2+\mu_y^2+c_1)(\sigma_x^2+\sigma_y^2+c_2)} \quad (2)$$

$c_3$ is usually equal to $c_2/2$. $\mu_x$ and $\mu_y$ are the mean values of x and y of the sample image respectively. $\sigma_x^2$ and $\sigma_y^2$ are the variances of x and y. $\sigma_{xy}$ is the covariance of x and y. $c_1$, $c_2$ are two constants, they keep the denominator from being zero.

### 4.2 Experimental results and analysis

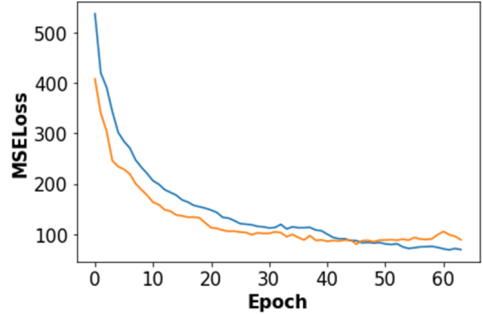

**Fig. 5.** The loss of training set and validation loss: blue is training loss and orange is validation loss.

**Fig. 5** above shows the loss of the 0-64 epoch training set and validation set, where blue is training loss and orange is validation loss. The training duration is 821.1742s. Finally, for images with the size of 1200*1200, the average loss of training set is 64.4132, and the average loss of validation set is 82.2553. The output images of training set, validation set and test set are shown in **Fig. 6**. The test set and validation set output results are shown in the figure below.

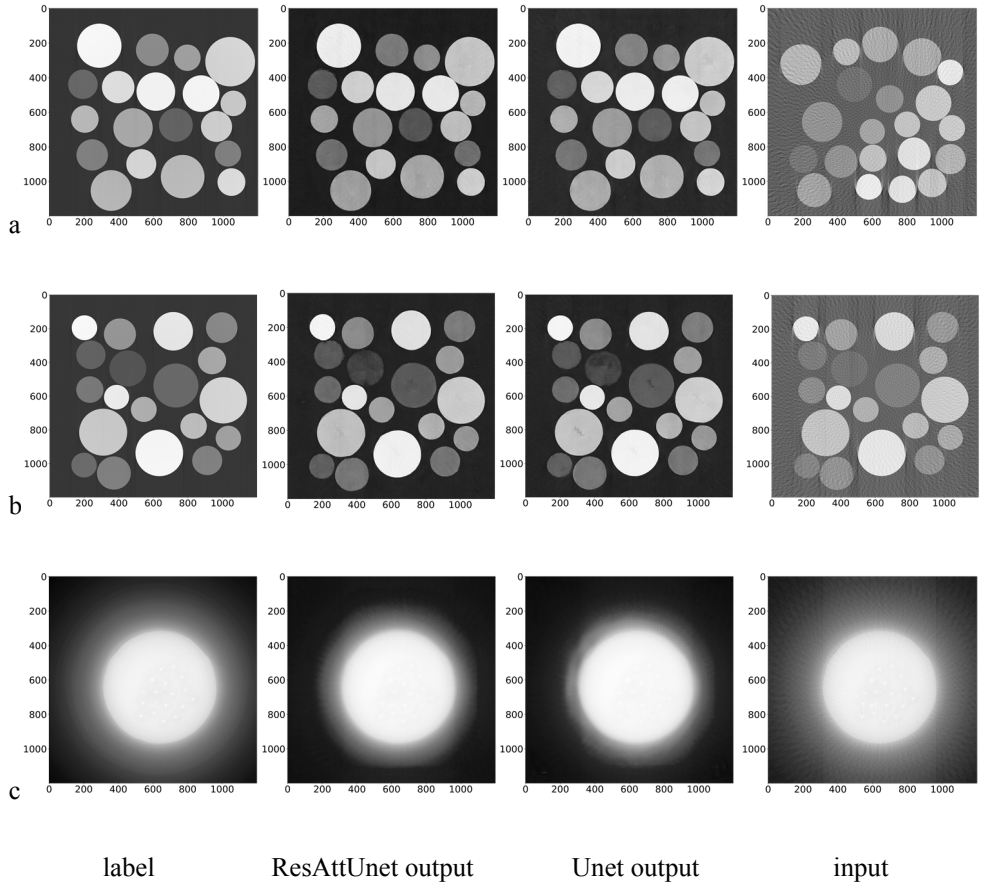

    label          ResAttUnet output       Unet output          input

**Fig. 6.** The output of ResAttUnet and Unet

    **Fig. 6** above shows label, ResAttUnet output, Unet output and data input from left to right, and a, b and c respectively represent the output results of training set, validation set and test set. According to the image results, there is not much difference between ResAttUnet and Unet on the training set. In the validation set, the output image of ResAttUnet model has some incomplete peripheral circles, while the output image of Unet has obvious holes in the inner circles. In the real data test set, both ResAttUnet and Unet have roughly removed

artifacts, but the small round holes in the output image of Unet are larger than the label. We used all the data from the training set, validation set and test set to conduct PSNR and SSIM evaluation, and averaged the evaluation values of all images to get the following table. O0 and O2 denote the opt level. O2 is the opt level used in our experiment, and O0 as the control group indicates that mixed precision training is not used. As can be seen from the data in the Table 1, various evaluation values of ResAttUnet are higher than those of Unet. And the mixed precision training has little effect on the results. Table 2 shows the reduction of video memory occupation and training time by using mixed precision training for different models.

**Table 1.** PSNR and SSIM of models

| Model | PSNR | | | SSIM | | |
|---|---|---|---|---|---|---|
| | Training set | Validation set | Test set | Training set | Validation set | Test set |
| **ResAttUnet(O0)** | 30.1283 | 29.0519 | 26.7286 | 0.9913 | 0.9891 | 0.9881 |
| **ResAttUnet(O2)** | 30.6647 | 29.0860 | 25.3447 | 0.9923 | 0.9894 | 0.9842 |
| **Unet(O2)** | 30.1168 | 29.0257 | 24.6563 | 0.9913 | 0.9890 | 0.9825 |

**Table 2.** Effects of mixed precision training on video memory occupancy and training time

| Model | O0 | | O2 | | | |
|---|---|---|---|---|---|---|
| | video memory (MiB) | training time(s) | video memory (MiB) | training time(s) | video memory saving | training time saving |
| **ResAttUnet** | 7021 | 1401.1953 | 4773 | 809.8071 | 32.01% | 42.20% |
| **Unet** | 5575 | 953.6296 | 4451 | 611.7596 | 20.16% | 35.84% |

From the output results of the model, the proposed model is effective. Compared with the traditional iterative method, it greatly reduces the time of image reconstruction. In addition, it also greatly reduces the workload of the experimenters. This work in this paper has a positive role in promoting the field of synchrotron radiation source CT. However, compared with Unet, the ResAttUent proposed in this paper still uses convolutional neural networks. Because the size of the convolution kernel is fixed, convolutional neural networks can only pay relative attention to local information. Using transformer or MLP might be better for handling global information.

## 5 Conclusion

In this paper, we propose a deep learning-based network model ResAttUnet for low-dose synchrotron radiation CT reconstruction. ResAttUnet adds the structure similar to ResNet and attention mechanism on the basis of Unet. The proposed method is tested on simulated and real datasets and compared with the Unet model. Experimental results show that the ResAttUnet proposed in this paper is effective in low-dose synchrotron radiation CT reconstruction. The use of mixed precision can reduce video memory requirements and training time.

## Reference

1. Zeng, G. . (2010). *Medical image reconstruction.*
2. Ronneberger, O. , Fischer, P. , & Brox, T. . (2015). U-net: convolutional networks for biomedical image segmentation.
3. K. H. Jin, M. T. McCann, E. Froustey and M. Unser, "Deep Convolutional Neural Network for Inverse Problems in Imaging," in IEEE Transactions on Image Processing, vol. 26, no. 9, pp. 4509-4522, Sept. 2017
4. Liu, Z. , Bicer, T. , Kettimuthu, R. , Gursoy, D. , & Foster, I. . (2020). Tomogan: low-dose synchrotron x-ray tomographywith generative adversarial networks. Journal of the Optical Society of America A, 37(3).
5. K. He, X. Zhang, S. Ren and J. Sun, "Deep Residual Learning for Image Recognition," 2016 IEEE Conference on Computer Vision and Pattern Recognition (CVPR), Las Vegas, NV, USA, 2016, pp. 770-778, doi: 10.1109/CVPR.2016.90.
6. Chaudhari, S. , Polatkan, G. , Ramanath, R. , & Mithal, V. . (2019). An attentive survey of attention models.
7. Vaswani, A. , Shazeer, N. , Parmar, N. , Uszkoreit, J. , Jones, L. , & Gomez, A. N. , et al. (2017). Attention is all you need. arXiv.
8. Gürsoy, Doğa, De Carlo, F. , Xiao, X. , & Jacobsen, C. . (2014). Tomopy: a framework for the analysis of synchrotron tomographic data. Journal of Synchrotron Radiation, 21(5), 1188-1193.
9. Xiaoming, Jiang, Esheng, Tang, Dingchang, & Xian. (1995). Beijing synchrotron radiation facility. Review of Scientific Instruments.